\documentclass[prl,twocolumn,showpacs,floatfix]{revtex4}

\usepackage{graphicx}

\begin{document}

\title{Quantum-critical behavior of itinerant ferromagnets}
\author{Andrey V. Chubukov$^1$, Catherine P\'epin$^2$ and
 Jerome Rech$^2$.}
\vspace{5mm}
\affiliation{$^1$Department of Physics, University of Wisconsin-Madison,
1150 University Ave., Madison WI 53706-1390\\
$^2$ SPhT, CEA-Saclay, L'Orme des Merisiers, 91191 Gif-sur-Yvette,
France }
\date{\today}
\begin{abstract}
We study the stability of the Quantum Critical Point (QCP)
 for itinerant ferromagnets commonly described by the
Hertz-Millis-Moriya (HMM) theory.  We argue that in $D \leq 3$,
 long-range spatial correlations
 associated with the Landau damping of the order parameter field
 generate a universal {\it negative}, non-analytic  $|q|^{(D+1)/2}$
contribution to the static magnetic
susceptibility $\chi_s (q, 0)$, which makes HMM theory unstable.
We argue that the actual transition is either towards incommensurate
 ordering, or first order.
We also show that singular corrections are specific to the spin problem,
while charge susceptibility remains
 analytic at criticality.
\end{abstract}

\maketitle

\vspace{5mm}

The critical behavior of itinerant paramagnets at the onset of
ferromagnetic ordering is the subject of intensive
experimental~\cite{exp} and theoretical~\cite{theory,dirk}
studies.
 Of particular interest is the existence of a
ferromagnetic second order transition at low temperatures, and
 the emergence of superconductivity near the phase boundary.
The critical theory of itinerant
 ferromagnets was developed by Hertz,  Millis, and Moriya~\cite{hertz} (HMM)
and is based on the action of the form:
\begin{equation}
S = \int d^d q d \omega_m
\left [ \xi^{-2} + q^2 + \frac{|\omega_m|}{|q|} \right]\phi^2_{q,\omega}
 + b_4 \phi^4 + b_6 \phi^6 + ...
\label{new1}
\end{equation}
where $\phi$ is a bosonic field associated with the order
paramater, and $\xi$ is
 the correlation length, which diverges at the phase transition.
  The two assumptions behind Eq.
(\ref{new1}) are that the
 $b_{2 n, n \geq 2}$ terms are non-singular and can be approximated by constants,
 and that the static spin susceptibility
 has  a regular  $q^2$ momentum  dependence.
The HMM theory has been successfully applied to explain
quantum-critical behavior in a number of materials~\cite{exp}, however
 its key assumptions have been questioned in recent
studies~\cite{abanov_1,bkv}. The analyticity of the $b_{2 n}$
terms was analyzed in detail for
 an anti-ferromagnetic transition where it was demonstrated that,
 for $D \leq 3$, all $b_{2n}$
pre-factors do have non-analytic pieces which depend on the ratio
between momenta and frequencies of the $\phi$
fields~\cite{abanov_1}. For a 2D anti-ferromagnetic quantum
critical theory (2D fermions and 2D spin fluctuations) these
non-analytic terms give rise to singular vertex corrections
~\cite{abanov_1}. For a 2D ferromagnetic case,
 non-analytic terms in $b_{2n}$ are still present, however we found that they
 do not give rise to anomalous exponent in the spin susceptibility for $D >1$
 and therefore are not dangerous.

In this communication, we question another key assumption of the
HMM theory, namely that  the momentum dependence of the static
spin propagator is analytic at small $q$. This assumption is based
on the belief that
 in itinerant ferromagnets
the $q$ dependence of the $\phi^2$ term comes solely from fermions
with high energies, of order of $E_F$, in which case
 the expansion in powers of
 $(q/p_F)^2$ should generally hold for $q \ll p_F$.
This reasoning was disputed in Refs.~\cite{bkv} and
~\cite{maslov}.
  These authors
 considered a static spin susceptibility $\chi_s (q)$
in a weakly interacting Fermi-liquid, i.e., well away from a quantum
 ferromagnetic transition,
 and argued that for $D \leq 3$ and arbitrary small interaction,
the small $q$ expansion
 of $\chi_s (q)$ begins with a non-analytic $|q|^{D-1}$ term.
 This  non-analyticity  originates from  a
 $2 p_F$ singularity in the particle-hole polarization
 bubble~\cite{bkv,maslov,millis} and
 comes from {\it low-energy} fermions
 with energies of the order of $ v_F q \ll E_F$.
  In ref.~\cite{bkv}, it was further argued
that within RPA, the non-analyticity in $\chi_s (q)$ gives rise to
the emergence of a non-analytic $|q|^{D-1} \phi_{q,\omega}^2$ term
in Eq. (\ref{new1}). Furthermore, the prefactor of this term turns
out to be negative, which signals the breakdown of the continuous
transition to ferromagnetism.

The weak point of this reasoning
 is that within RPA  one assumes that fermionic excitations remain
coherent at the QCP.
 Meanwhile, it is known~\cite{ioffe} that approaching the QCP,
 the fermionic
 effective mass $m^*$ diverges as $\log \xi $ in $D=3$ and as
  $\xi^{3-D}$ in smaller dimensions. Moreover, $m/m^*$ appears as a
prefactor of the $|q|^{D-1}$ term, hence the prefactor
 {\it vanishes} at the QCP.  Does this imply that
Eq. (\ref{new1}) is valid at the transition? Not necessarily,
since one has to verify explicitly whether or not the divergence
of $m^*$
 completely eliminates a non-analyticity  in $ \chi_s (q) $,
or just makes the non-analytic term weaker than away from QCP.
If the latter is true, the non-analytic term
 can still be much larger than $q^2$ at small $q$ at criticality.

In this communication, we report explicit calculations which show
that the non-analytic term
 is still present at the QCP and accounts for the breakdown
of the HMM theory. We first
 consider the problem in the Eliashberg approximation
  and argue that
 this approximation leads to  a non-Fermi liquid behavior at the QCP,
 but the magnetic susceptibility $\chi_s (q)$ remains analytic in $q$.
We then demonstrate that
 the corrections to the Eliashberg theory are singular and
 give rise a universal
 contribution to $\chi^{-1}_s (q)$ in the form
$q^\beta$, with $\beta \leq 2$ for $D \leq 3$ and a {\it negative}
pre-factor.

The starting point for our theory is a low-energy effective
spin-fermion Hamiltonian obtained by integrating the fermions with
energies between fermionic bandwidth $W$
 and $W \Lambda \ll W$ out of the partition function~\cite{ioffe,dirk}:
\begin{eqnarray}
{\bf H} &=& \sum_{k, \alpha} v_F (k -k_F) c^{\dagger}_{k,\alpha}
c_{k, \alpha} +
\sum_q \chi^{-1}_{s,0} (q) {\bf S}_q {\bf S}_{-q} \nonumber \\
&& + g \sum_{k,q} c^{\dagger}_{k+q,\alpha} {\bf \sigma}_{\alpha,\beta} c_{k,\beta} {\bf S}_q
\label{new2}
\end{eqnarray}
where ${\bf S}_q$ describe collective bosonic degrees of freedom in the spin
channel,  $g$ is the residual spin-fermion coupling, and
$\Lambda$ (the upper cutoff) is
 roughly the scale up to which fermionic dispersion can be linearized near the
 Fermi surface.
 The bare spin susceptibility $\chi_{s,0} (q)$
is assumed to be analytic at $q \ll p_F$ and have an
Ornstein-Zernike form  $\chi_{s,0} = \chi_0/(\xi^{-2} + q^2)$.
 The coupling and $\chi_0$ appear in the
 theory only in combination ${\bar g} = g^2 \chi_0$.

The perturbation theory for Eq. (\ref{new2})  is an
expansion into two parameters\cite{dirk}:
\begin{equation}
\begin{array}{lll}
  \alpha = \frac{3^{3/4}}{4 \pi}~\frac{\bar g}{E_F}&
  \mbox{and} &
  \lambda \equiv \frac{m^*}{m}-1  =  \frac{3}{4\pi}~ \frac{\bar g}
{v_F \xi^{-1}}~\sim \alpha ~(\xi p_F) \end{array}
  \end{equation}
where $E_F $ is the electron's Fermi
 energy and the pre-factor in $\alpha$ is chosen for future convenience.
We assume that ${\bar g}$ is small compared to $E_F$, i.e.,
$\alpha $ is a small parameter~\cite{dirk,gorkov}. That $g \ll E_F$
 is in line with the very idea of a
separation  between high-energy and low-energy physics.
 At the same time, near a
 ferromagnetic QCP, $\xi$ diverges, and the
 dimensionless coupling $\lambda \propto \xi$ is large.
Fortunately, one can solve explicitly
the strong coupling problem in $\lambda$ while keeping
 the zeroth order in $\alpha$~\cite{ioffe,dirk}.
 This amounts to neglecting the vertex correction $\delta g/g$ and the
  momentum dependent piece in the self-energy $\Sigma (p)$,
  which are of next order in $\alpha$.
The analogous theory for
 electron-phonon interaction is known as Eliashberg theory~\cite{eliash}.

{\it Eliashberg theory}~~
The Eliashberg theory for a ferromagnetic quantum critical point
 has been described in detail in the
literature~\cite{ioffe,dirk}. The results are presented here for
$D=2$, but their validity holds for general $D leq 3$.
 The elements of the theory are
coupled fermionic and bosonic self-energies $\Sigma (\omega_m)$ and
$\Pi (q, \omega_m) = \Phi (\omega_m)/|q|$, respectively~\cite{dirk}
($G^{-1} (p, \omega_m) = i(\omega_m + \Sigma (\omega_m)) - v_F (p-p_F);~
\chi_s (q, \omega_m) = \chi_0/(q^2 + \xi^{-2} + \Pi (q, \omega_m))$).
The self-consistent solution yields
$\Phi (\omega_m) =\gamma |\omega_m|$,
 where $\gamma = {\bar g} p_F/(\pi v^2_F)$, and
\begin{equation}
\Sigma (\omega_m) = \lambda \omega_m~
f\left( \frac{(p_F \xi)^3 \omega_m \bar{g}}{E_F^2}\right),~~~
\label{new3}
\end{equation}
 where $f(0) =1, ~f(x \gg 1) = (32 \pi/3\sqrt{3}x)^{1/3}$,
and  $E_F = p_F v_F/2$.
 At  the QCP, $\xi = \infty$
and $\Sigma (\omega_m) = \omega_m^{2/3} \omega^{1/3}_0$
 where $\omega_0 = \alpha^2 E_F$ is the typical bosonic frequency of
 our problem. Note that $\omega_0 \ll E_F$.
The $\omega^{2/3}$ dependence of $\Sigma$ implies that at the QCP,
the Fermi liquid description is  broken down to
 the lowest energies. At the same time, Eliashberg theory reproduces
 the form of the spin propagator from Eq. (\ref{new1}),
$ \chi_s (q,\omega_m) \propto (\xi^{-2} + q^2 + \gamma |\omega_m|/q)^{-1}$.
 Alternatively speaking, in the Eliashberg approximation,
 the Fermi liquid is destroyed at QCP, but
 the magnetic transition remains continuous,
 and $\chi_s (q)$ is analytic.

\begin{figure}[tbp]
\centerline{\includegraphics[width=2.7in]{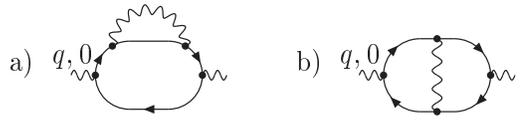}}
\caption{
The two lowest-order diagrams for the corrections to the static spin susceptibility. The self-energy insertion is $\Sigma (p)$, the frequency dependent piece $\Sigma (\omega_m)$ is already incorporated into ``zero-order'' Eliashberg theory}
\label{fig1}
\end{figure}

{\it Beyond Eliashberg theory}~~
The validity of the Eliashberg theory is generally based on Migdal theorem
 that states that vertex correction $\delta g/g$ and
$\Sigma (p)$ are small when the bosonic velocity $v_B$
is much smaler than $v_F$. In our case,  a'posterioiri
 analysis of typical bosonic and fermionic momenta which contribute
 to $\Phi (\omega)$ and $\Sigma (\omega)$ within the Eliashberg theory
shows that for the same
 $\omega \sim \omega_0$, typical $q_B  \sim (\gamma \omega_0)^{1/3}
\sim \alpha p_F$, while  typical fermionic
momenta  $|k - k_F| \sim \omega_0 /v_F \sim \alpha^2 p_F$
 are much smaller. This implies that for $\alpha \ll 1$, the
 effective bosonic velocity   $v_B \sim \alpha  v_F$
 is much smaller than  $v_F$, i.e., Migdal theorem is valid.
To verify this, we
 computed the  corrections to  $\partial \Sigma (p) /\partial p$
 and to the static vertex  in the limit of zero bosonic momentum and
 for external fermions at the Fermi surface and at
zero frequency,  and  indeed found $\alpha^{1/2}$ smallness.

Still, this does not immediately imply that $\chi_s (q)$ remains
 analytic beyond Eliashberg theory.  To verify this,
 we have to go beyond the limit of zero momentum, evaluate
the correction to the static susceptibility at
 a finite $q$ and check whether it
 remains analytic in $q$. Let's do this.
Consider e.g., the diagram  in Fig. \ref{fig1}a) with the
 insertion of $\Sigma (p)$ into the particle-hole bubble.
 Expanding this diagram to order $q^2$
 we find that it scales as  $q^2 I$, where
\begin{equation}
I \propto \alpha^{2} \frac{E_F^{3}}{\gamma^{1/3}}
 \int \frac{d l_x~d \Omega |\Omega|^{2/3}}{(v_F l_x - i |\Omega + \Sigma(\Omega)|)^5}  S\left(\frac{l^2_x}{(|\Omega| \gamma)^{2/3}}\right)
\label{n1}
\end{equation}
where $l^2 = l^2_x + l^2_y$,  $S(x) \propto \int dl_y /(l^2 +
|\Omega| \gamma/|l|)$ normalized to $S(0) =O(1)$ and $S(x \gg 1) \propto
x^{-1/2}$.  The integrand for $I$  has a highly degenerate pole in the upper
half-plane of $l_x$, at $v_F l_x \propto |\Omega + \Sigma (\Omega)|$.
 This degenerate
 pole can be avoided by closing the integration contour
in the lower half-plane of $l_x$, in which case the non-vanishing
part in $I$ comes from the non-analyticity in $S(x)$ which,
 we recall, is the  boson propagator
integrated  along the Fermi surface. This physically implies that fermions
 undergo forced vibrations at typical bosonic frequencies.
 The issue then is what are the typical bosonic $x$ for the non-analyticity
in $S(x)$. By analogy  with the electron-phonon problem, one could expect
that corrections to the Eliashberg theory come from the processes in which
  typical bosonic frequencies are near bosonic mass shell.
Then typical $l_x$ are of order $(\Omega \gamma)^{1/3}$ as in the Eliashberg theory, i.e., typical $x$ in $S(x)$ are of order one.
Substituting this typical $x$ into (\ref{n1}), we obtain $I \sim
\alpha^{1/2}$, i.e. small and analytic correction to the static
 susceptibility $\delta \chi_s (q) \propto q^2 \alpha^{1`/2}$.
 However, a careful examination of the
scaling function $S(x)$
 reveals that it is  {\it non-analytic} already at the smallest
 $x$, such that there is another contribution to $I$
from bosons vibrating at typical $l_x$ comparable to those near
 {\it fermionic} mass shell.
Indeed, expanding $S(x)$ in powers of $x$ at $x \ll 1$ we obtain
\begin{equation}
S(x) = 1 - \frac{3\sqrt{3}}{8\pi}~
\frac{l_x^2}{(\gamma |\Omega|)^{2/3}}~ \log [l^2_x]
\label{new6}
\end{equation}
The $\log l^2_x$ term  is the most important here - its presence implies that
$S(x)$ possesses a branch cut along the  imaginary axis of $l_x$ down to the
 smallest $l_x$.  One can easily make sure that this
 logarithmic singularity is the
 consequence of the $\Omega/|l|$ non-analyticity of the Landau damping piece
in the bosonic propagator,
 i,.e., of the existence of the
 long-range spatial component of the dynamical susceptibility.
Substituting  the small $x$ form of $S(x)$ into (\ref{n1}),
 and evaluating integrals we find a much larger, divergent
 contribution to $I$
  which  behaves as $ |q|^{-1/2}$. This implies that the low-$q$ expansion of
 $\delta \chi_s$ is actually non-analytic in $q$ and begins as $q^{3/2}$.
Performing  an explicit calculation without expanding in
$q$,  and combining the results from the two
diagrams for $\chi (q)$ in Fig. \ref{fig1} we  obtain for
the static spin susceptibility
\begin{equation}
\chi_{s} (q) = \frac{\chi_0}{q^2 - 0.17 |q|^{3/2} p^{1/2}_F}
\label{n2}
\end{equation}
Observe that the $q^{3/2}$ term  in (\ref{n2}) is not small in $\alpha$.
  This is not surprising,
 since the only excitations involved are those near the fermion mass shell,
 while $\alpha$
 measures how soft are the mass shell bosons
 compared to the mass shell fermions.
Note in passing that the self-energy insertion in the diagram of Fig.
\ref{fig1}a) is not a double counting, since the correction to
 static susceptibility only comes
 from  the momentum-dependent piece in the self-energy.
The frequency-dependent $\Sigma (\omega)$ is already incorporated at the
``zero-order'' level, and it does not affect $\chi_s (q)$.

The non-analyticity of the correction to the susceptibility and the
 disappearance of  $\alpha$ could be also detected by
 analyzing vertex corrections away from  the limit of $q=0$
 and zero fermion frequency $\omega$.
Indeed, evaluating $\delta g(q,\omega)/g$ at
finite $q$ and $\omega$, we obtain
\begin{equation}
\delta g (q,\omega) \propto g  \alpha^{1/2}~
\psi \left(\frac{m{\bar g}}{q^2},\frac{\omega}{E_F}\right)
\label{new4}
\end{equation}
 where $\psi (x \ll 1,y \ll 1) \propto \sqrt{x} \ll 1$, and
$\psi (x \gg 1,y \ll 1) =
1 + A log y/\sqrt{x}$, $A = O(1)$. We see that, although
$\psi (x,y) \leq 1$ (i.e.
vertex correction is small in $\alpha^{1/2}$),
the small $q$
expansion of $\delta g(q)/g$ begins
 as $\alpha^{1/2} |q||\log \omega|/\sqrt{m{\bar g}}
\sim |q| |\log \omega|/p_F$, i.e., it is non-analytic in $q$ and in $\omega$,
 and the $|q \log \omega|$ term does
 not contain $\alpha^{1/2}$.  This singular term again comes from fermions
near their own resonance and can be traced back to the non-analyticity of the Landau damping term. Substituting this $\delta g(q,\omega)$
into the susceptibility diagram
 and performing computations we find that $\log \omega$ dependence
 makes $\delta \chi_s (q)$ nonzero in the static limit (without it, the correction would be proportional to $\Phi (\omega)$ and vanish at $\omega=0$),
and the
 overall power of $q$ is $|q|$ from the vertex correction
times the ratio of typical $\omega$ and $v_F l_x$.
As typical $v_F q_x \sim \Sigma (\omega) \propto \omega^{2/3}$, this ratio yields $\omega^{1/3} \propto (l_x)^{1/2} \sim q^{1/2}$, hence the overall power in
 $\delta \chi_s (q)$ is $q^{3/2}$, as above.

 What happens with higher-order terms? We analyzed higher-order
 particle-hole insertions into the susceptibility bubble
 and found that they add only small,  $O(\alpha^{1/2})$,
 corrections to the $q^{3/2}$ term in (\ref{n2})
 since at least one internal momentum is near the boson mass
 shell. Higher-order particle-particle (cooperon)
insertions give rise to
 $O(1)$ corrections  to the $q^{3/2}$ term
and in principle should be included. However, the non-smallness
 of particle-particle insertions is not specific to our problem
 and is in fact  customary in
Eliashberg-type
theories~\cite{theory,eliash,abanov_1,csp,dirk,gorkov}.
Particle-particle vertex corrections are known to  give rise to
 a pairing instability close to the ferromagnetic QCP at $T_{c} = O(\omega_0)$,
 which
 implies that there exists a dome around the QCP where the normal state analysis
 is invalid. However, a ferromagnetic $T_c$ has an order of
 magnitude of roughly
$T_{c} \leq 0.015 \ {\omega_0}$~\cite{dirk}. Hence the typical
bosonic momenta for the pairing
 are of order $0.01 \alpha p_F$,
much smaller than typical $q$ in Eq. (\ref{n2}). Due to this
separation of energy scales,
 we do not expect particle-particle
 insertions to be relevant.

Eq. (\ref{n2}) is the central result of the paper. It shows that
the static
 susceptibility becomes negative around a ferromagnetic QCP at
 $q < q_1 =0.029 \ p_F$.
Although the pre-factor is  small,  $q_1$ is parametrically larger
than the typical fermion and boson momenta that contribute to the
fermion self-energy in the quantum-critical regime
 ($|k-k_F| \sim \alpha^2 p_F$ and $q_B \sim \alpha p_F$, respectively).
 This implies that the whole quantum-critical region
  is in fact {\it unstable}.
Alternatively speaking, $Z=3$ quantum-critical behavior at a ferromagnetic
 transition is internally destroyed by infrared singularities
 associated with the
 non-analytic momentum dependence of the Landau damping.

A negative static susceptibility up to a
 finite $q$ implies either that the instability occurs
when $\xi^{-1}$ is still finite, into an incommensurate state with
a finite $|q| \sim q_1$~\cite{braz}, or the ordering
 is commensurate, but  the magnetic transition
 is of the first order. Which of the two scenarios holds remains to be studied.
We also verified that away from the QCP, when $\xi$ is finite, the
$q^{3/2}$ term
 transforms into $|q| \xi^{-1}$ at the smallest $q$. The interpolation formula
 between the $q^{3/2}$ and $|q|$ forms is rather involved and we will not
 present it here.

For completeness, we also computed the temperature variation of the uniform
$\chi_{s} (q=0)$ at the QCP and in 2D obtained, with logarithmic accuracy,
\begin{equation}
\chi_s (T) = - \frac{2 m {\bar g}}{3\pi} \left(\frac{T}{E_F}\right)~
\log\left[\frac{E_F}{T}\right]
\label{n3}
\end{equation}
That $\chi_s (T, q=0)$ is {\it negative}
 is another indication that the system is unstable at the QCP.

 For arbitrary $D<3$, our calculations yield the
``correction'' to $\chi_{s}^{-1}$
 in the form $|q|^{(D+1)/2}$, again with a negative prefactor.
In 3D  it scales as  $q^2 \log |q|$.
Hence, HMM theory is  destroyed for all $D\leq 3$. For
$D>3$, the correction term is smaller than $q^2$, and HMM theory
survives.

{\it Charge susceptibility and gauge theory.} $~~~$
We repeated the same calculations for
 the charge susceptibility and found in D=2, that
 the singular $|q|^{3/2}$ terms from the two diagrams in Fig. \ref{fig1}
cancel each other, and the $q^2$ behavior survives. The same holds
for the propagator of the gauge field -  singular $|q|^{3/2}$ terms from
 self-energy and vertex correction insertions into the particle-hole bubble
 again cancel each other. We verified the cancellation of non-analytic terms
 for the second-order diagrams as well.
Furthermore, we verified in two orders of perturbation
that the momentum dependence of $\chi_c (q, 0)$
 comes in powers of $(q/p_F)^2$, i.e., it is entirely  due to  the existence of a curvature of the dispersion.
 Physically, the  distinction
 between spin and charge susceptibilities is in that
  the $\Omega/|q|$ singularity  in the dynamical particle-hole response
function (which generates  the $q^{3/2}$ non-analyticity),
 is sensitive to a magnetic field, but
 is insensitive to the
 change of the chemical potential. Hence, the singularity
 shows up in the total spin response, but  does not appear in the total charge
 response~\cite{kirk}.
Mathematically, this distinction is due to the presence, in the spin case,
 of the Pauli matrices in the vertices of the diagrams in Fig. \ref{fig1},
 such that the self-energy and vertex correction diagrams
 do not cancel each other.
 Note  that the cancellation or non-cancellation of the
singularities is not  directly related to the conservation laws
for charge and spin susceptibilities, since conservation laws
 only require that $\chi_{c} (q=0, \omega)$ and $\chi_{s} (q =0, \omega)$
 vanish, but impose no constraints on the forms of the susceptibilities in the
 other limit of finite $q$ and zero frequency.
More specifically, Fermi-liquid relations between
 $\chi_{s,c} (q \rightarrow 0,\omega =0)$ and
$\chi_{s,c} (q=0, \omega \rightarrow 0)$ imply~\cite{landau}
 that
$\chi_{s,c} (q\rightarrow 0, \omega =0) \rightarrow Cst$, but
impose no
 formal constraints on  the
 form of the actual $q-$ dependence in $\chi_{s,c} (q, \omega =0)$ at
 finite $q$.

{\it Conclusion}~~ To summarize, we studied the stability of the
quantum critical point for
 itinerant ferromagnets, in the limit when the spin-fermion coupling
is much smaller than the bandwidth, and
 one would naively expect Eliashberg theory to be valid.
Within Eliashberg theory, the Fermi liquid is destroyed at
criticality, but
 the magnetic  transition is still continuous, and
 described by bosonic Hertz-Millis-Moriya action.  We demonstrated, however,
 that in  $D \leq 3$,
 long-range spatial correlations
 associated with the Landau damping of the bosonic order parameter
 field,
break the Eliashberg theory and
 give rise to a universal, {\it negative}, non-analytic  $|q|^{(D+1)/2}$
contribution to the static magnetic susceptibility $\chi_s (q,
0)$. This term makes the continuous critical theory unstable. We
argued that the actual transition is either towards incommensurate
ordering, or first order into a commensurate state. We also
demonstrated that singular corrections are specific to the spin
problem, while  charge susceptibility remains
 analytic at criticality.

It is our pleasure to thank  A. Abanov, I. Aleiner, A. Andreev, B.
 Altshuler, D. Belitz, P. Coleman, S. Das Sarma, A. Georges, L. Glazman,
 A. Finkelstein, Y.B. Kim, T. Kirkpatrick, D. Maslov, A. Millis, M.
 Norman, O. Parcollet, I. Paul,
 R. Ramazashvili  and Q. Si for stimulating discussions.
The research has been supported by NSF DMR 9979749 (A. V.Ch.).

\end{document}